\documentclass[]{article}

\usepackage{graphicx}
\usepackage{subcaption}
\usepackage{amsfonts}
\usepackage{amsmath}
\usepackage{float}

\usepackage[
    backend=biber,
    style=nature,
  ]{biblatex}
\title{Semi-Classical Cutoff Energies for Electron Emission and Scattering at Field-Enhancing Nanostructures with Large Ponderomotive Amplitudes}
\author{Joshua Mann, James Rosenzweig\\
University of California, Los Angeles}

\begin{document}

\maketitle

\begin{abstract}
	 The uniform field assumption used to derive semi-classical cutoff energies of $10U_p$ for electron emission and $3.17U_p$ for high harmonic generation is applicable for ponderomotive amplitudes ($\propto E\lambda^2$) much smaller than the field drop-off scale. For large wavelength and high field experiments at nanoscale structures this assumption may break down by predicting energies beyond the true classical energy limits. Here we provide generalized calculations for these cutoff energies by taking into account the spatial field drop-off. The modified cutoff energies vary significantly from the uniform field results even with ponderomotive amplitudes still an order of magnitude below the field drop-off scale. Electron emission and scattering energy as a function of the time-of-ionization is considered for the nanotip ($\sim1/r^2$) field profile. The cutoff energies as a function of the adiabaticity parameter $\delta$, which may be easily calculated for given wavelength, apex field strength, and nanostructure scale, are then determined through maximization for nanotip, nanoblade ($\sim1/r$), and exponential field profiles. These profiles deviate from each other in electron emission energy by up to nearly a factor of the ponderomotive energy, indicating the importance of mid-field profile behavior. The electron emission energy cutoff also attains an additional factor of $U_p$ due to the smooth integrated ponderomotive force in the adiabatic drop-off and very long pulse regime. These results also provided as double-exponential fits for ease of use. We then compare the nanoblade electron emission cutoffs with a quantum simulation of the electron rescattering process. We also consider a short (few-cycle) pulsed field, focusing on a cosine-like pulse and overviewing the general carrier-envelope phase dependencies.
\end{abstract}

\section{Introduction}

The study of strong-field physics is extensive, involving both gaseous \cite{gas1nhhg, gas2nhhg, gas3nhhg} and solid-state \cite{bigrev, mat1, mat2, mat3, mat4nhhg, mat5} systems. The underlying processes of electron elastic scattering result in high harmonic generation (HHG) \cite{gas1nhhg, gas2nhhg, gas3nhhg, mat4nhhg} in addition to high energy electron emissions (EE) \cite{mat1, mat2, mat3, mat5, ntip1, ntip2}. The interaction between intense laser fields and solid nanostructures permits peak ionization field strengths in the GV/m range with sub-wavelength confinement \cite{bigrev, ntip1, ntip2} allowing for more intense and localized emissions. These benefits are particularly attractive, although the solid-state systems involve the complexities of many body effects \cite{mat3,manydmg1} (including the limitations of commonly utilized single active electron models of HHG \cite{singlim}), surface roughness \cite{manydmg2}, and the potential for target damage \cite{bigrev, mat3}. Nanotip electron sources provide spatially coherent beams, albeit with low intensities \cite{bigrev} largely due to this damage threshold. One way to increase yield is by utilizing an array of nanotips \cite{array2, array1}. Extending the emission area from a nanotip to a nanoblade may also provide more electrons due to its larger area and may allow stronger surface fields due to its improved thermomechanical properties \cite{myself, gerard, timo}. In the strong-field regime, and particularly for gases, the pseudo-free electron dynamics dominate much of the phenomena observed.

The semi-classically derived cutoff energies for EE and the corresponding HHG emission, both stemming from the underlying process of electron rescattering, are commonly quoted as being $10U_p$ and $3.17U_p$ respectively. These values assume a spatially uniform field, which holds true when the field profile is unchanging on the scales of the ponderomotive amplitude $a_p$. Solid-laser interaction experiments typically utilize the field enhancement properties of nanometric structures (such as nanotips), where the field drops off on some length scale $R$ (radius of tip). As $a_p$ approaches this length scale the assumption of uniform field breaks down. Another work has investigated this process at specific parameters for an exponential field profile \cite{otherpap}. It was shown that the quiver motion is significantly altered for sufficiently small decay lengths at a given wavelength and field strength, resulting in a reduced peak energy. In this manuscript we provide the generalized calculations at these structures, ultimately yielding modified cutoff energies for both EE and HHG (electron scattering, ES) in terms of more translatable quantities such as the structure scale, ponderomotive amplitude, ponderomotive energy, and finally the adiabaticity parameter all in the system's natural units.

Our ES calculations do not include the electron hole dynamics \cite{otherpap} or any particularities of the band structure \cite{classhhg} which are highly relevant for solid-based HHG. Their inclusion would introduce a separate set of natural units, with field decay length and effective mass (or dispersion relation) dependent on the material, and so a more thorough study would need to be done specifically for each unique system.

\section{Classical Model}

To derive a classical limit to the emission and scattering energies involved in electron rescattering we will consider a semi-classical model of the process. The only quantum component here is the appearance of the electron at some time-of-ionization (TOI) $t_0$ with zero initial energy at the material surface, attributed to an ionization process such as above-threshold ionization (ATI) (ignoring excess energy) or quantum tunneling (ignoring the tunneling distance). The classical model used to show that the peak energies for electron rescattering are proportional to the ponderomotive energy, $U_p=\frac{e^2E^2}{4m\omega^2}$, is simply the kinematic equation in an oscillating field

\begin{equation}
\ddot{x}(t)=\frac{eE}{m}\cos\left(\omega t\right)
\end{equation}
\begin{equation}
\dot{x}(t_0)=x(t_0)=0
\end{equation}

Additionally we must include elastic scattering at the surface. To do this, whenever the electron strikes the surface, we simply reverse the velocity. Solving this problem numerically until the end of one cosine period ($t_f=2\pi/\omega$) will yield the final energy for a given TOI $t_0$. Emitted electrons would continue to quiver in the field, but at this termination time there is no contribution of this quiver motion to the kinetic energy, hence why it is chosen. The quiver energy decreases to zero once the electron sufficiently escapes the system or the pulse terminates. Maximizing this emission energy with respect to $t_0$ will then provide the $\approx 10U_p$ EE cutoff. One may perform the same maximization method, but instead maximize with respect to the scattering velocity, to yield the HHG/ES cutoff of about $3.17U_p$. Both cutoffs also attain an extra quantum mechanical factor \cite{bigrev} which we will not consider here without loss of utility.

Experiments that involve this process typically study, if not gases, nanotips. These nanoscopic structures induce enhanced electric field profiles that may taper off quickly compared to the ponderomotive amplitude, or quiver amplitude, of the field: $a_p=\frac{eE}{m\omega^2}$. This is particularly applicable for large wavelength experiments (small $\omega$). Once this regime is approached, the model utilizing a uniform laser field breaks down. Instead, our system now follows a differential equation of the form

\begin{equation}
\ddot{x}=\cos(t)g(x)
\end{equation}

where $g(x)$ is the spatial profile of the field and we have renormalized into the natural units of this system. $x$ is now in units of the peak ponderomotive amplitude $a_p$ and $t$ is in units of $1/\omega$, inverse of the laser frequency (effectively radians). In this way the field drop-off may be directly related to the adiabaticity parameter \cite{otherpap} by 

\begin{equation}
\delta=-\left(\frac{dg}{dx}(0)\right)^{-1}
\end{equation}

$\delta \gg 1$ indicates that the field is roughly constant and so dynamics are unaffected while $\delta \ll 1$ indicates the field gradient is strong enough to alter the dynamics of the system. In Ref \cite{otherpap} $\delta$ was defined using the $1/e$ decay length of the field, but here we generalized by using a gradient which is identical for an exponential field profile. This preferentially encapsulates the near-field behavior of the profile which is much more important for moderate ponderomotive amplitudes, $\delta\sim10$, and particularly so for pulsed driving lasers.

The simulation will run from $-\frac{\pi}{2}<t_0<\frac{\pi}{2}$ to $t_f=2n\pi$. The electron's kinetic energy is given by $2U_p\dot{x}^2$, and so the modified ponderomotive constant is $\chi=2\dot{x}^2$ with $\dot{x}$ maximized against $t_0$ at either the end of the simulation (EE energy) or at the time of scattering (HHG/ES energy). The ponderomotive values $U_p$ and $a_p$ are representative of the field's properties at $x=0$, such that $g(0)=1$, which we will here take to be the peak field achieved.

In the uniform field model the final energy was taken to be the energy at time $2\pi$, once the electron has experienced one full laser period. However, now that our field is not uniform, the final emitted energy may further depend on other factors. For instance, if we have a field drop-off that corresponds to some ponderomotive force $F_p=-\nabla U_p(x)$, one may expect that the final energy of the electron is not simply its kinetic energy after one full period. Instead, once the electron has fully left the system, it will have experienced a ponderomotive force corresponding approximately to an additional emission energy of $U_p$. So, one may adjust the ponderomotive multiplier to include an additional factor representing the remaining ponderomotive energy. Here we will run with 10 laser cycles ($t_f=20\pi$) to ensure that the electron has sufficiently interacted with the field. We will then add the integrated final ponderomotive force $E_{corr}=U_p g^2(x_f)$. For large $\delta$ the electron remains in the near-field for the duration of the simulation and so this factor should be of order $E_{corr}\approx U_p g^2(0)=U_p$. For small $\delta$ the electron will have traversed most of the field profile and so this factor should be $E_{corr}\approx U_p g^2(\infty)=0$.

Experiments where this system usually applies oftentimes use short (few-cycle) pulses and so they may hit a temporal termination before the electron experiences the field out to infinity. To address this, we additionally consider a Gaussian-pulsed laser for some cases. Otherwise, we will opt to assume a continuous laser field where the electron will experience the entire ponderomotive force, and this final quantity is what will be maximized. This assumption is roughly applicable for $\tau\gg R\sqrt\frac{m}{20U_p}$. Fortunately for lower energy experiments with short pulse lengths, where the opposite extreme is approached, the standard uniform field results would be applicable.

\section{Numerical Methods}

Here we describe the methods used for this semi-classical simulation. The individual particle simulation is performed as described in Equations 2 and 3 by breaking up Equation 3 into two first-order equations. The first equation, for position, follows Heun's method (which is equivalent to a second order Taylor series in this case)

\begin{equation}
x'=x+h_0v+\frac{h_0^2}{2}\cos(t)g(x)
\end{equation}

with $h_0$ being the time step size, chosen to be $10^{-3}$ for this study. The velocity follows Heun's method as well, however utilizing this new position instead of the Euler method approximation for its next-step evaluation. This substitution reduces the amount of operations required and slightly improves the accuracy (but the method's order remains the same)

\begin{equation}
v' = v + \frac{h_0}{2}\left(\cos(t)g(x)+\cos(t')g(x')\right)
\label{eq6}
\end{equation}

with $t'=t+h_0$. We also have our initial condition at $t=t_0$ of $x=0$ and $v=0$ from Equation 2.

These iterations apply for when the electron is not bound to hit the surface during the current time interval. To account for this infrequent but essential case we check for a series of conditions.

If the electron is bound to be in the material, $x\neq0$ and $x'\leq0$, then we need to elastically scatter at the surface. To do this, we first need to determine the time step required to put the electron exactly at the surface. If our acceleration at this time step is identically zero, then our new time step will be have to be $h_s=-\frac{x}{v}$. Otherwise, the time step is given by $h_s=\left(-|v|+\frac{v}{|v|}\sqrt{v^2+2ax}\right)/a$. This may pose issues numerically if the acceleration is very small, but we have not encountered problems stemming from this. A Taylor series in $a$ may be performed to get around this issue. We then set the position $x=0$, calculate the new velocity according to Equation \ref{eq6} with $h_0\rightarrow h_s$, and then reverse the velocity. If one wishes to find the ES velocity for the HHG cutoff then they could terminate the simulation here.

If we are within one time step of the end of the simulation, $2n\pi-t<h_0$, then we run with a modified time step $h_f=2n\pi-t$ in the same manner as normal. This ensures that we do not accrue any extraneous energy which would occur if we stop the simulation offset from the end of the cosine cycle.

Lastly, if the other cases are not met, then the simulation runs as normal. Once the end is reached, we have a final velocity (or a scattering velocity) that may be maximized against $t_0$.

\subsection{Determining Maximum Energy}

To determine the maximum energy we must equivalently maximize the final velocity (whether it be the scattering velocity for HHG/ES or the velocity at $t=2n\pi$ for EE). We take a relatively straightforward approach to this to find the global maximum.

Since these simulations run fairly quickly, we start with a fine sweep of 100 samples evenly spread among the domain of $t_0\in\left[-\frac{\pi}{2}, \frac{\pi}{2}\right]$. We then take the $t_0$ associated with the maximal velocity, $t_c$, and perform two more simulations at times $t_0=t_c\pm h_0$ and proceed with Newton's method. We connect a parabola to these three points in $v$ vs $t_0$ space, and set our new maximal time $t_c$ to be at the vertex of the parabola (if the concavity is non-negative, the step size should be reduced or the initial sample number should be increased). This process then repeats until the change in peak velocity between steps is smaller than a threshold, here taken to be $10^{-9}$. As long as the initial sample size is fine enough a global maximum should be reached.

\section{Uniform Field Profile}

The standard uniform field calculation assumes that the laser-electron interaction terminates once the pulse slowly ends temporally and adiabatically, and the laser field is assumed to have no spatial dependence. In this light, we do not include the extra ponderomotive force contribution to the final velocity as there is none in a uniform field.

\begin{figure}[H]
	\centering
	\includegraphics[width=1\linewidth]{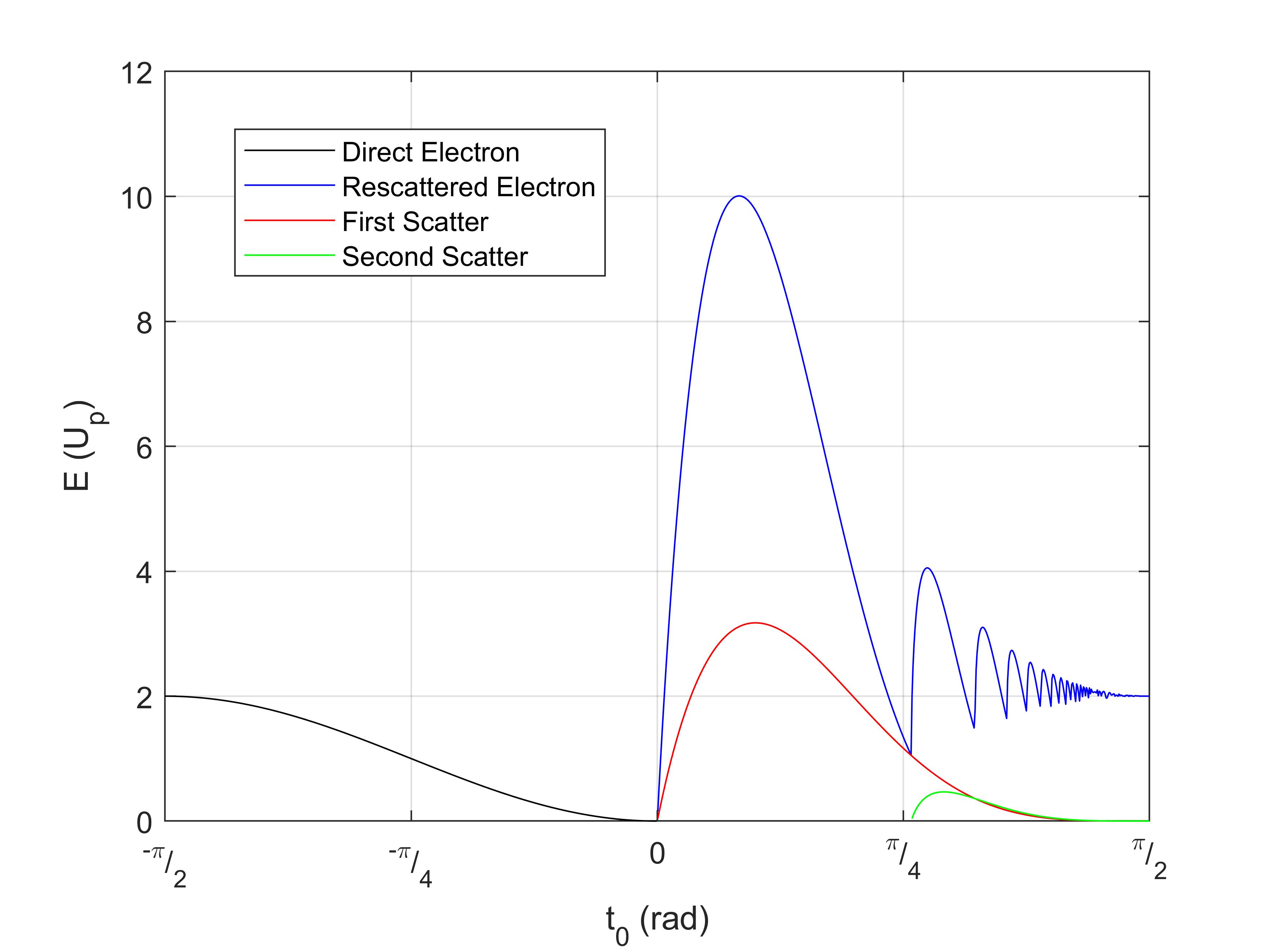}
	\caption{The emission and scattering energy dependencies on $t_0$ for a uniform field. Direct electron emission (black) has a clear cutoff at $2U_p$. At $t_0=0$, scattering begins to occur, and the rescattered electron energy (blue) approaches a peak energy before undergoing multiple scattering events. The subsequent scattering events do not lead to any higher energies. The first (red) and second (green) scattering energies are also shown, the former leading to the HHG classical cutoff.}
\end{figure}

The relation between the emission phase $t_0$ and the emission and scattering energies, which we will now refer to as the emission energy landscape (EEL), is shown in Figure 1. The direct electron emission spectrum has a classical cutoff at $2U_p$ (which can be derived analytically). At $t_0=0$ we begin to see scattering events, with only the first providing a boost in energy. 

Performing the aforementioned maximization techniques provide the commonly quoted semi-classical cutoff for EE of $10.0076U_p$ at $t_0=0.2610\approx\pi/12$. Given that this result is usually applied to quantum-natured spectra, which obfuscates the cutoff, such precision is not typically required which is why the $10U_p$ rounding is common and useful. The ES cutoff is $3.1731U_p$ at $t_0=0.3134\approx\pi/10$. This result is more commonly rounded as $3.17 U_p$.

\section{Non-uniform Field Profiles}

Contrary to the uniform field profile calculations, with a non-uniform field it is generally unclear as to whether the laser-electron interaction terminates due to a temporal boundary or a spatial boundary. A laser pulse of sufficiently short length would end the interaction before the electron traverses the entire field profile, and so the pulse length and the near- to mid-field behavior would have the largest impact on results. Alternatively, a long pulse would permit the electron to explore the most of the field profile and so then the field profile itself would be the only independent variable. For this reason we continue assuming an infinitely long pulse, or $\tau\gg R\sqrt\frac{m}{20U_p}$.

Because of this, we will append the remaining ponderomotive energy at the end of the calculation as mentioned before. This gives rise to a potentially confusing result -- the cutoff energy for electron emission with large adiabaticity parameter would then be $\sim 11 U_p$ instead of $\sim 10 U_p$ due to the integrated ponderomotive force. The only way to reconcile this issue is by specifying a particular pulse length or temporal profile, which we will do in Section 6. Otherwise, this additional factor of $U_p$ is expected in experiments with extremely long pulse lengths (according to the assumption above) and large $\delta$.

\subsection{Nanotip-like Fields}

The enhanced field profile from a nanotip will follow a $1/r^2$-type drop-off. The most accurate and neatly analytic form of the field would be $E_0\left[(\gamma-1)\left(\frac{R}{R+x}\right)^2+1\right]$ with $\gamma$ being the enhancement factor, $R$ the tip radius, $x$ the distance from the tip surface, and $E_0$ the unenhanced field ($\gamma E_0$ is what would be used to calculate $U_p$ and $a_p$). This model includes the feature that the field enhancement factor drops off to unity instead of zero. This necessitates the free parameter $\gamma$, however, and so we will opt to use a simpler field in this scenario. The field profile we will use is

\begin{equation}
g(x) = \left(\frac{R}{R+x}\right)^2=\left(\frac{2\delta}{2\delta+x/a_p}\right)^2
\end{equation}

The general field behavior is retained here while the extra dimension of the enhancement factor is lost. This is effectively the limit of large enhancement factor with $\gamma$ being absorbed into the peak field strength. The adiabaticity parameter is then $\delta=\frac{R}{2a_p}$.

\begin{figure}[H]
	\centering
	\includegraphics[width=1\linewidth]{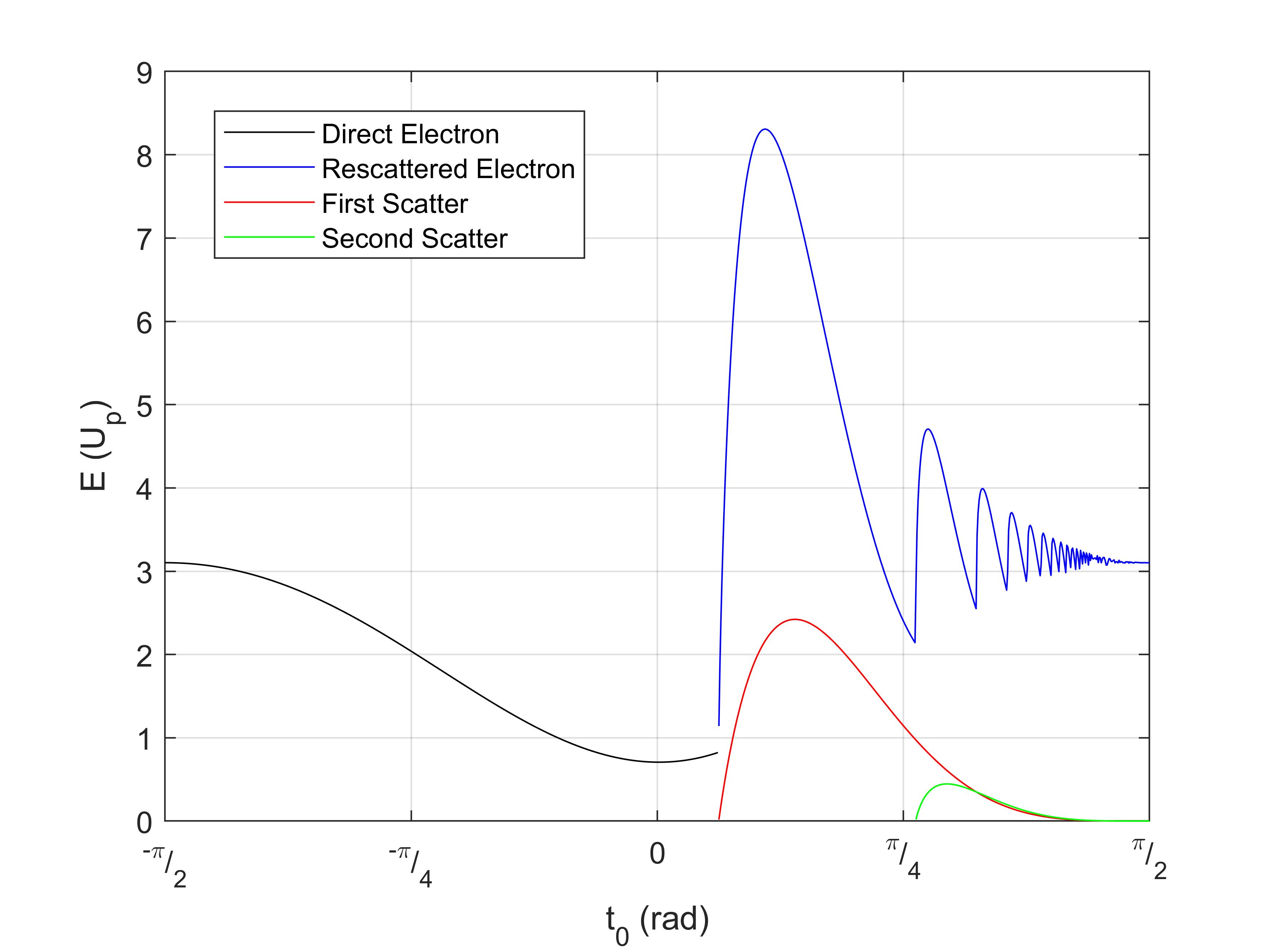}
	\caption{EEL for a nanotip-like field profile, $a_p=0.1R$ or $\delta=5$.}
\end{figure}

A slowly decaying field profile, with ponderomotive amplitude only $a_p=0.1 R$, $\delta=5$, yields an EEL shown in Figure 2. The major differences are that: the direct electron cutoff has been raised from $2U_p$ to about $3U_p$, the ER maximum has been quenched to about $8U_p$ and to a later TOI, and the scatter energy has been lowered to about $2.5U_p$. The gap between direct and rescattered energies in these plots is due to the sample grid, even with 1000 points uniform across the $\pi$-sized domain.

\begin{figure}[H]
	\centering
	\includegraphics[width=1\linewidth]{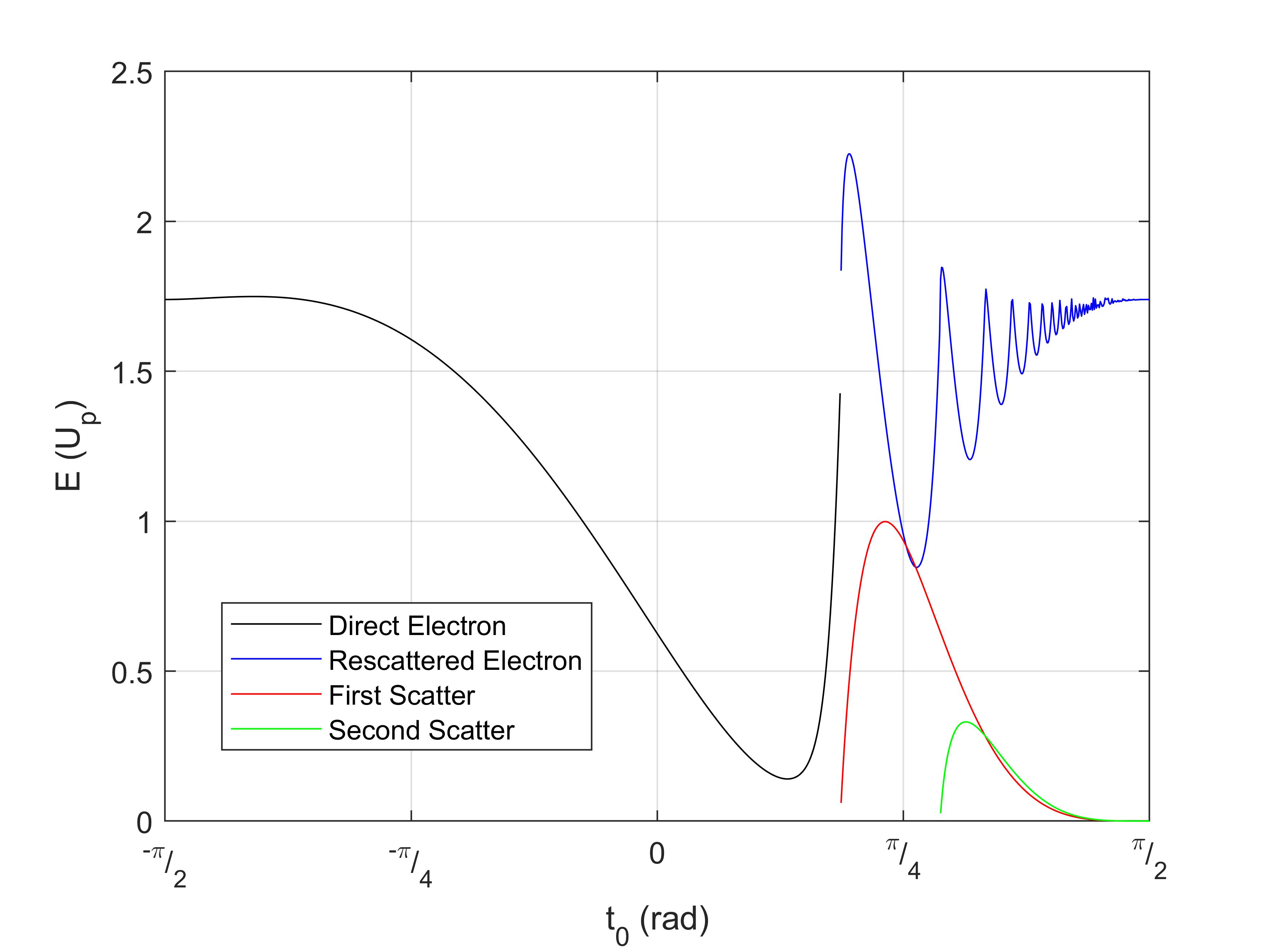}
	\caption{EEL for a nanotip-like field profile, $a_p=R$ or $\delta=\frac{1}{2}$.}
\end{figure}

Increasing the ponderomotive amplitude to $a_p=R$, $\delta=1/2$, yields an extremely altered landscape shown in Figure 3. It can be seen that the scattering event does not contribute much more energy to the peak electron emission when compared to the direct emission. The widespread reduction in energy is attributed to the drastic field dropoff. The electron quickly traverses into the mid-field for most TOIs. If the electron is emitted towards the end of the ionization half-period, the electron does not travel very far into the field and the restoring half-period is enough to still cause a low energy scattering event.

\begin{figure}[H]
	\centering
	\includegraphics[width=1\linewidth]{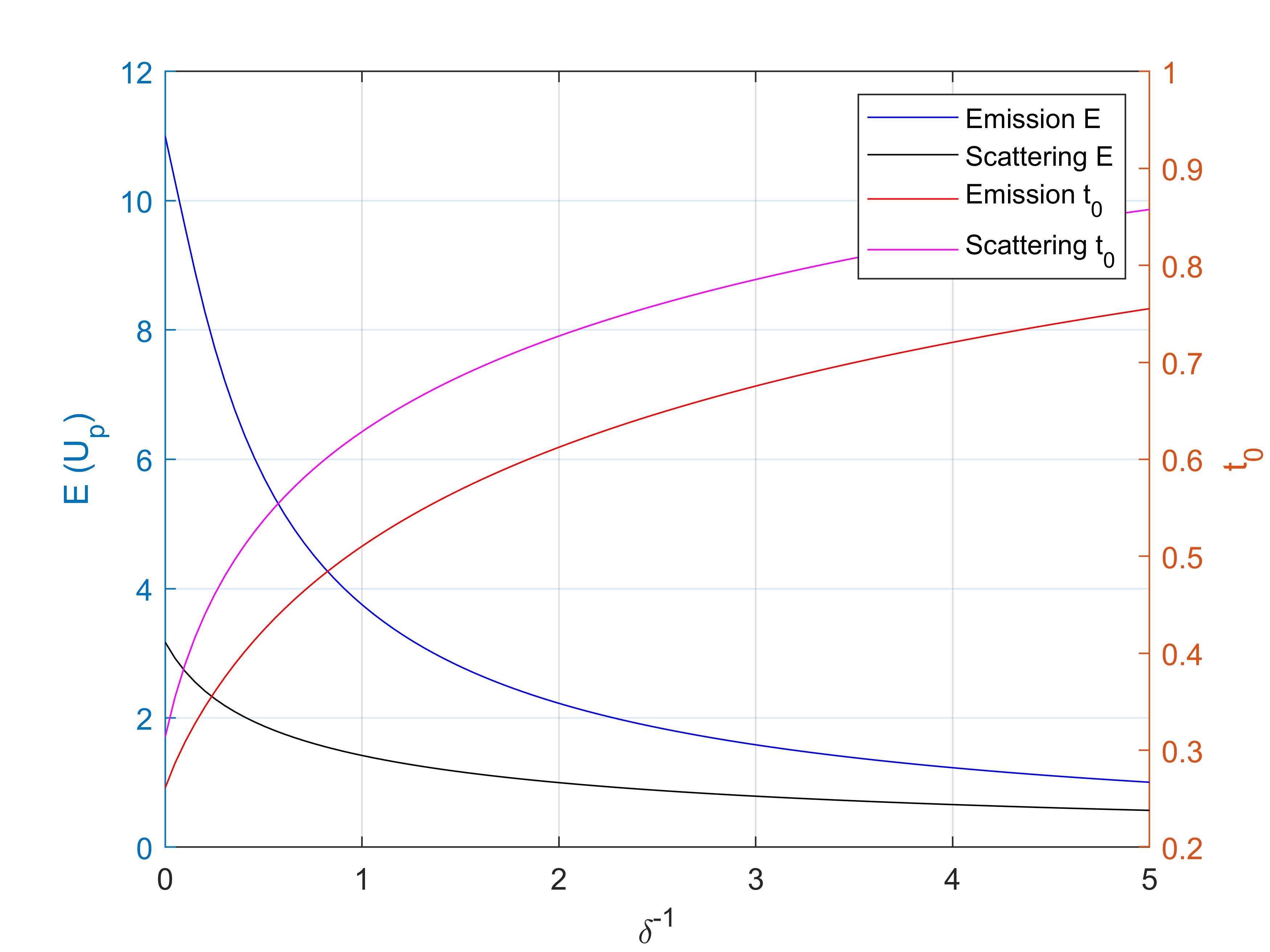}
	\caption{Cutoff behavior for tip fields. Emission (blue) and scattering (black) cutoff energies (left axis) are functions of the adiabaticity parameter. Additionally, the time of ionization (right axis) that leads to this peak energy is plotted for the emission (red) and scattering (magenta) cutoffs.}
\end{figure}

Performing the maximization technique for EE and ES provides the relationship shown in Figure 4. The cutoff energy monotonically decreases as the ponderomotive amplitude increases and the adiabaticity parameter decreases. 

As the ponderomotive energy scales differently from the ponderomotive amplitude with $E_0$ and $\lambda$, there is no single peak energy for this system. Instead, this depends on the specific electric field and wavelength. As $U_p/a_p\propto E_0$, one may keep $a_p$ fixed while increasing $E_0$ (decreasing $\lambda$), which then proportionately increases the peak energy emitted. Alternatively, if one wishes to have the smallest energy spread, they may do the opposite and work with high wavelengths and moderate fields. This is clearly moving in the direction of static field emission which of course does not involve a rescattering process.

The peak emission TOI monotonically increases, further departing from the peak ionization field at $t_0=0$. This may lead to fewer electrons being part of the spectral plateau. This behavior may also be seen in Figures 1-3 where the increase in ponderomotive amplitude pushes the direct electron emission into the positive $t_0$ region, decreasing the total number of electrons that go through the rescattering process.

The cutoff dependency on the adiabaticity parameter may be well estimated in this region using a double-exponential fit

\begin{equation}
\frac{E_{emit}}{U_p} \approx 7.352e^{-2.079/\delta}+3.656e^{-0.2729/\delta}
\end{equation}

\begin{equation}
\frac{E_{scat}}{U_p} \approx 1.658e^{-2.365/\delta}+1.515e^{-0.209/\delta}
\end{equation}

With root-mean-square error (RMSE) 0.040 and 0.025 respectively. The fits enforce the uniform field results for $\delta\rightarrow\infty$, including the additional factor of $1U_p$ for EE from the integrated ponderomotive force. The near-uniform truncated series expansions, taken by fitting a parabola exactly to three sample points near $a_p/R=0$, $\delta\rightarrow\infty$, are

\begin{equation}
\frac{E_{emit}}{U_p} \approx 11.0076-14.0166\delta^{-1}+0.5517\delta^{-2}
\end{equation}

\begin{equation}
\frac{E_{scat}}{U_p} \approx 3.1731 - 5.4743\delta^{-1} + 9.9577 \delta^{-2}
\end{equation}

The evaluation points are taken as $a_p/R=0, 0.025, $ and $0.05$.

\subsection{Nanoblade-like Fields}

Nanoblade fields are very similar to nanotips in their general form, however with one dimension removed the decay rate is reduced. The general analytic form is $E_0\left[(\gamma-1)\left(\frac{R}{R+x}\right)+1\right]$, and we will perform the same simplification to remove the enhancement factor

\begin{equation}
g(x) = \frac{R}{R+x}=\frac{\delta}{\delta+x/a_p}
\end{equation}

Here the adiabaticity parameter is $\delta=\frac{R}{a_p}$, double that of the nanotip for the same edge radius. This indicates that these quenching effects are weakened for the blade even while the same intense enhanced fields, and therefore ionization rates, may be reached.

\begin{figure}[H]
	\centering
	\includegraphics[width=1\linewidth]{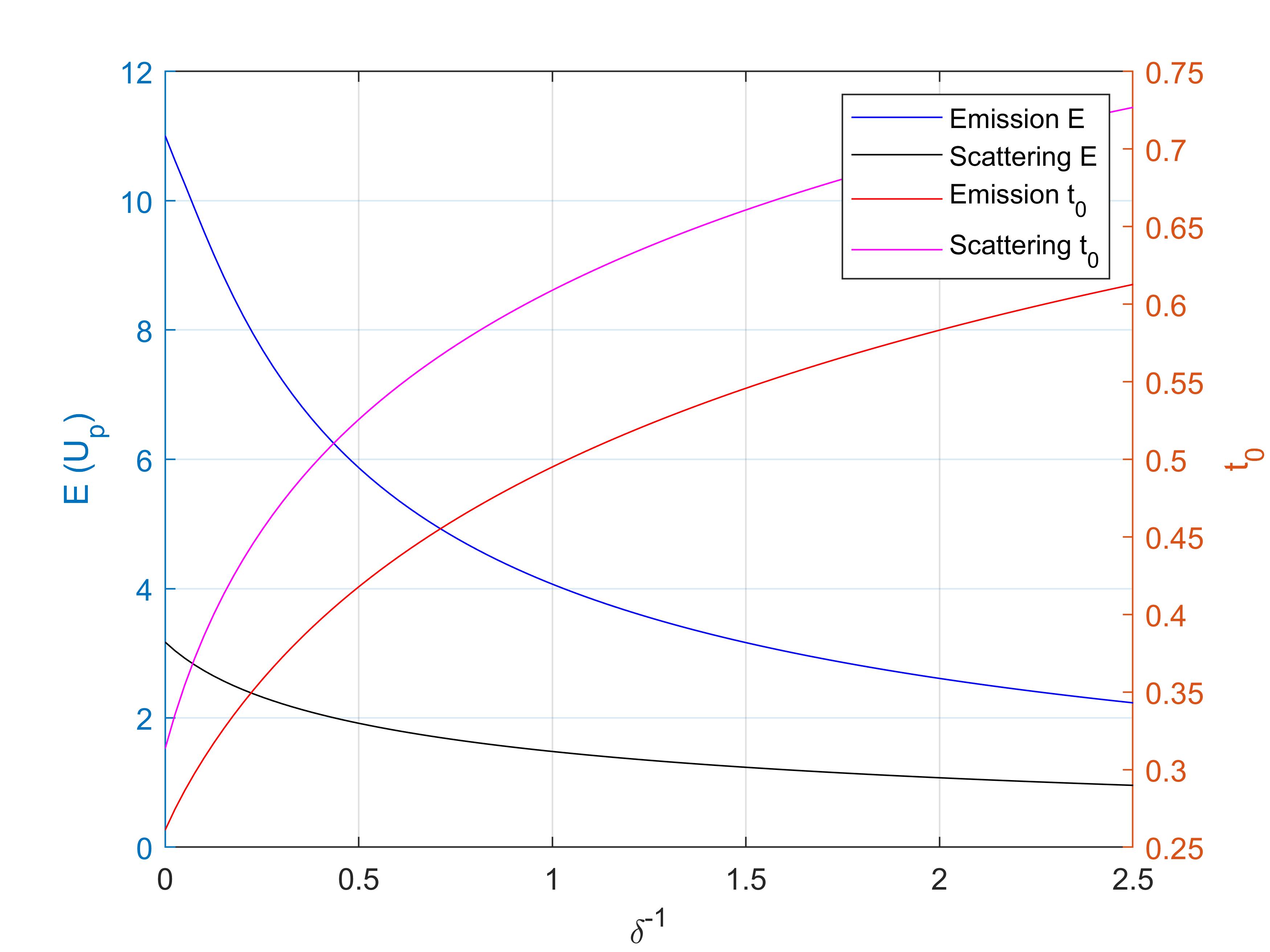}
	\caption{Cutoff behavior for nanoblade fields. Emission (blue) and scattering (black) cutoff energies (left axis) are functions of $\delta$. Additionally, the time of ionization (right axis) that leads to this peak energy is plotted for the emission (red) and scattering (magenta) cutoffs.}
	\label{fig5}
\end{figure}

For small ponderomotive amplitudes this field decays at half the rate of the tip field. The emission and scattering cutoff energies, as well as the peak emission time, are shown in Figure \ref{fig5}. Unsurprisingly we see the same effects as with the tip profile. Double-exponential fits for the emission and scattering energies are

\begin{equation}
\frac{E_{emit}}{U_p} \approx 5.952e^{-2.623/\delta}+5.056e^{-0.3344/\delta}
\end{equation}

\begin{equation}
\frac{E_{scat}}{U_p} \approx 1.285e^{-3.178/\delta}+1.888e^{-0.2817/\delta}
\end{equation}

with RMSE of 0.022 and 0.014 respectively. The near-uniform truncated expansions, in the same manner as before, are

\begin{equation}
\frac{E_{emit}}{U_p} \approx 11.0076 - 15.4369 \delta^{-1} + 13.0467 \delta^{-2}
\end{equation}

\begin{equation}
\frac{E_{scat}}{U_p} \approx 3.1731 - 5.5761 \delta^{-1} + 12.9153 \delta^{-2}
\end{equation}

\subsubsection{Comparison to TDSE Results}

\begin{figure}[H]
	\centering
	\includegraphics[width=1\linewidth]{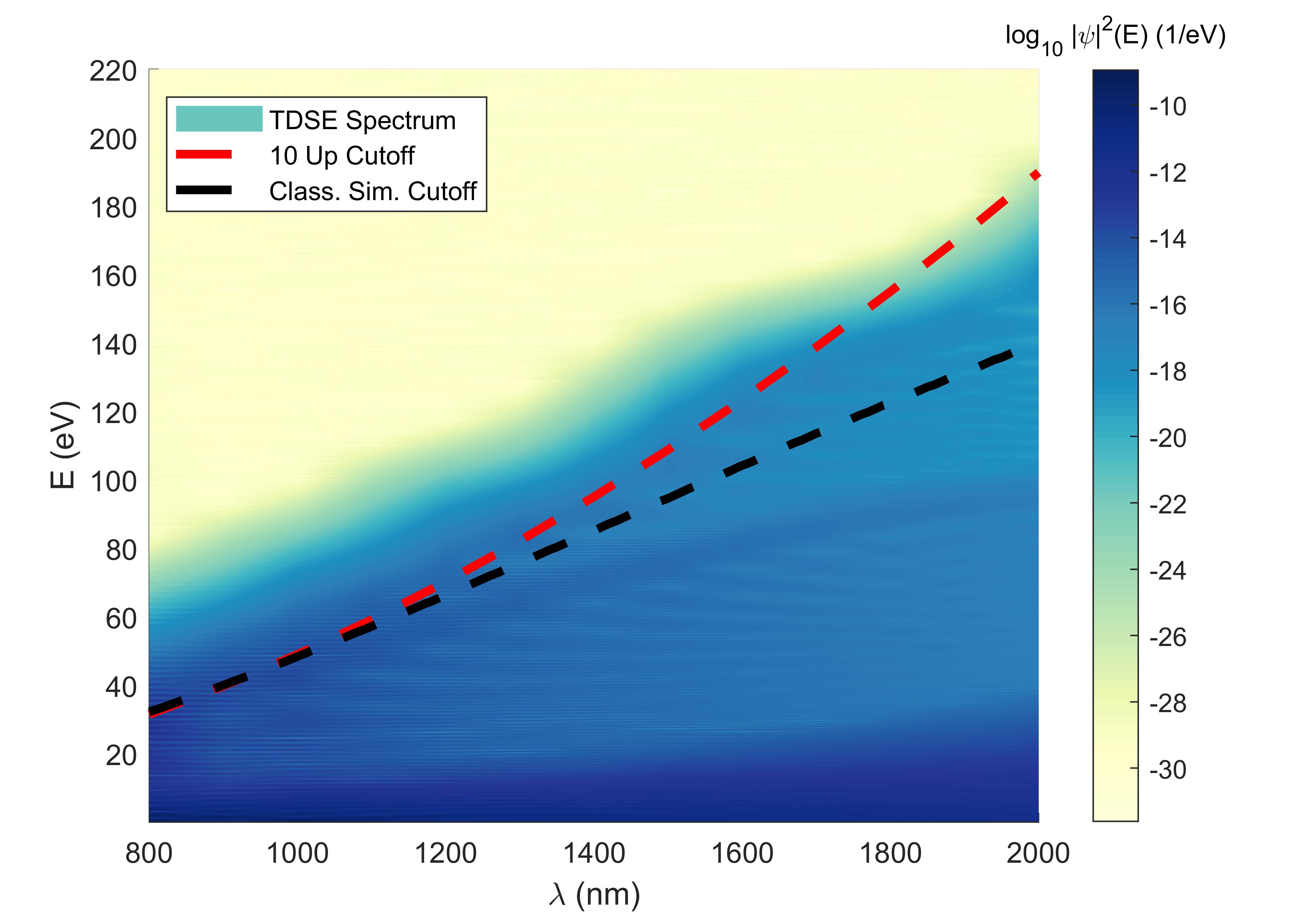}
	\caption{TDSE-derived spectra (colormap) with the semi-classical $10U_p$ cutoff (red dashed) along with the cutoff calculated in this section (black dashed). The TDSE spectra were calculated using a cylindrically decaying field with a peak field strength of 20 GV/m.}
	\label{fig6}
\end{figure}

In Ref. \cite{myself} we calculated the evolution of a Jellium surface-bound wave function under the influence of a strong laser field via the time-dependent Schr\"odinger equation (TDSE) and found the emitted electron spectrum. A strong deviation between the simulated spectra and the $10U_p$ cutoff arose for high wavelengths, even with the change in effective wavelength and peak field from the pulse window taken into account. Figure \ref{fig6} shows the comparison between the standard $10U_p$ approximation and the results found here overlaid on the TDSE spectrum. The TDSE spectrum's cutoff appears to behave roughly linearly with wavelength, coinciding well with our results, whereas the $10U_p$ cutoff behaves quadratically.

\subsection{Exponential Drop-off}

A common simple model of the field drop-off models the profile as an exponential function of the form

\begin{equation}
g(x)=e^{-x/\delta}
\end{equation}

for a given adiabaticity parameter ($x$ in units of $a_p$). This accurately models the near-field drop-off while ignoring the somewhat less important mid- and far-field effects of the particular geometry.

Double-exponential fits for the emission and scattering energies are

\begin{equation}
\frac{E_{emit}}{U_p} \approx 8.849e^{-1.719/\delta}+2.159e^{-0.1907/\delta}
\end{equation}

\begin{equation}
\frac{E_{scat}}{U_p} \approx 1.366e^{-3.105/\delta}+1.807e^{-0.03369/\delta}
\end{equation}

with RMSE of 0.061 and 0.013 respectively. The near-uniform truncated expansions are

\begin{equation}
\frac{E_{emit}}{U_p} \approx 11.0076 - 14.7772 \delta^{-1} + 36.7627 \delta^{-2}
\end{equation}

\begin{equation}
\frac{E_{scat}}{U_p} \approx 3.1731 - 5.5934 \delta^{-1} + 11.3491 \delta^{-2}
\end{equation}

\subsection{Comparison}

\begin{figure}[H]
	\centering
	\includegraphics[width=1\linewidth]{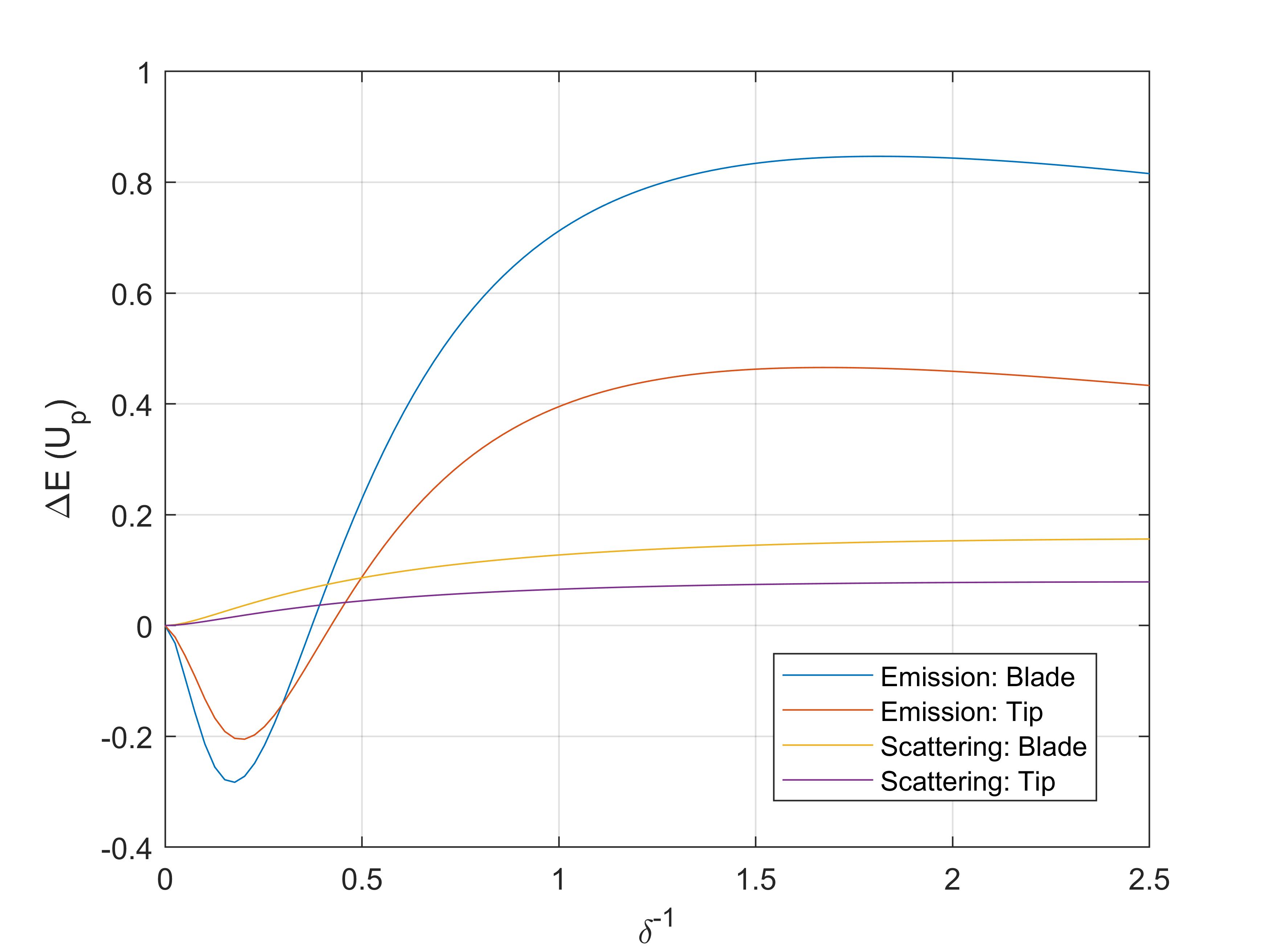}
	\caption{Deviation of emission and scattering energy for tip and blade fields from the exponential field drop-off.}
	\label{figcomp}
\end{figure}

A comparison of the emission and scattering energies for the tip and blade field, represented as a deviation from the exponential field, is shown in Figure \ref{figcomp}.

Beginning with emission, we see that for large adiabaticity parameter (near-uniform field) the geometry-specific cutoffs deviate negatively, whereas once $\delta$ decreases to $\sim 2-3$ we see a net growth in the energy, with diminishing returns quickly setting in. Since $\delta\rightarrow0$ indicates an infinitesimally short field we expect the deviation to tend towards zero, as with the energy for each profile, after this point. We note that, for the same adiabaticity parameter, the blade-like field profile decays spatially the slowest, then next is the tip profile, and finally the exponential field profile decays the fastest.

For large, but finite, $\delta$ the rescattering dynamics are relatively unaffected. However, $\delta$ is still small enough that the electron will experience vastly different field profiles within a even a single laser cycle. Considering the electron of maximal emission energy for the exponential field profile with $\delta^{-1}=0.3$, the field magnitude at the electron's position has already diminished to $0.02$ of its original strength after two laser cycles ($x\approx 13a_p$). The tip and blade profile magnitudes are reduced to 0.11 and 0.20 at this time, respectively. Because of this, the post-scattering exponential electron experiences a strong extraction force (about the same for all cases) while the following restoring force is greatly reduced (by about a factor of 3 compared to blade fields) permitting larger emission energies with the same scattering dynamics for the exponential field.

However, once $\delta^{-1}>0.5$ the scattering dynamics are meaningfully altered. The scattering time is pushed later and the scattering energy is reduced for the exponential field compared to the blade and tip fields. Not only is the emission energy for tip and blade fields larger due to the extra scattering energy, but they experience the second extraction field for longer (because of the earlier scattering time) and gain more energy from the field (because of the larger scattering velocity).

The increase in scattering energy for blade and tip fields compared to the exponential field has a relatively straightforward explanation: the fields are closer to uniformity because of the slower decay rate (including the nonlinear orders beyond $\delta$) and so the rescattering dynamics are closer to the ideal uniform system which attains a larger scattering energy.

\section{Gaussian Pulsed Cutoffs}

Typically a pulsed laser is used in ER and HHG experiments. The laser pulse envelope may be treated as a Gaussian, providing the spatio-temporally dependent field:

\begin{equation}
\frac{E(x,t)}{E_0} = g(x)e^{-2\ln 2 \left(\frac{t}{\tau}\right)^2}\cos(t-\phi)
\label{eq22}
\end{equation}

with $\tau$ being the full-width half-max power and $\phi$ the carrier-envelope phase (CEP), both in units of radians. We will focus on $\phi=0$ (electric field is strongest when ionizing, not restoring) and $\tau=6\pi$ (akin to an 800 nm, 8 fs pulse).

\begin{figure}[H]
	\centering
	\includegraphics[width=1\linewidth]{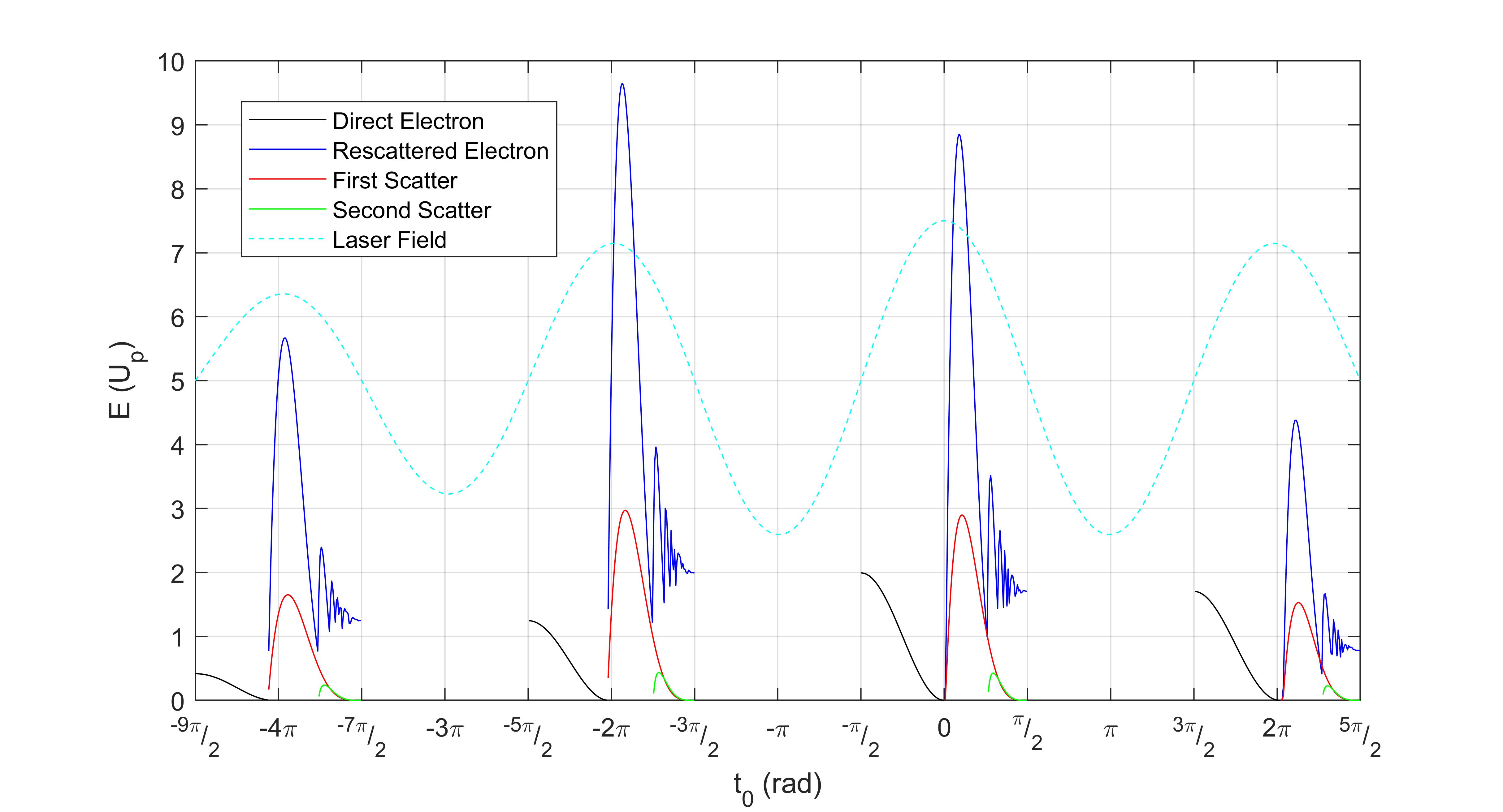}
	\caption{EEL for a pulsed uniform field, $\tau=6\pi$, $\phi=0$.}
	\label{fig7}
\end{figure}

The EEL for this laser pulse with a uniform spatial field profile is shown in Figure \ref{fig7}. We note that the direct electron max energy is still around $2U_p$, emitted at the center of the envelope. This is expected as the envelope is only slowly decaying post-emission, and the integrated field after the central cosine wave will roughly zero out.

The peak rescattered emission energy is a different story. The peak energy comes from electrons that were ionized during the cycle before the peak field, $t_0\sim-2\pi$. The majority of the energy gained by the maximal-energy electron occurs after the scattering event (scattering electron energy is about $3U_p$, final energy is about $10U_p$, so about $7U_p$ is gained after scattering). So, the field should be strongest after the electron scatters to maximize emission energy, which is why it must be ionized in the cycle before the peak cycle.

The peak scattering energy also originates from a $t_0$ during the cycle preceding the peak field. This is likely because the scattering energy is assisted by the fact that the restoring field is stronger than the ionizing field (as it is closer to the envelope peak), resulting in a pseudo-ponderomotive force brought by a field strength gradient in time. This excess restoring force results in a larger scattering energy, and thus a larger scattering cutoff than those of the emissions during the peak cycle. Although, the peak scattering energy originating from ionization at the peak field is not too much lower, likely because the restoring field is of the same strength for both cases.

\begin{figure}[H]
	\centering
	\includegraphics[width=1\linewidth]{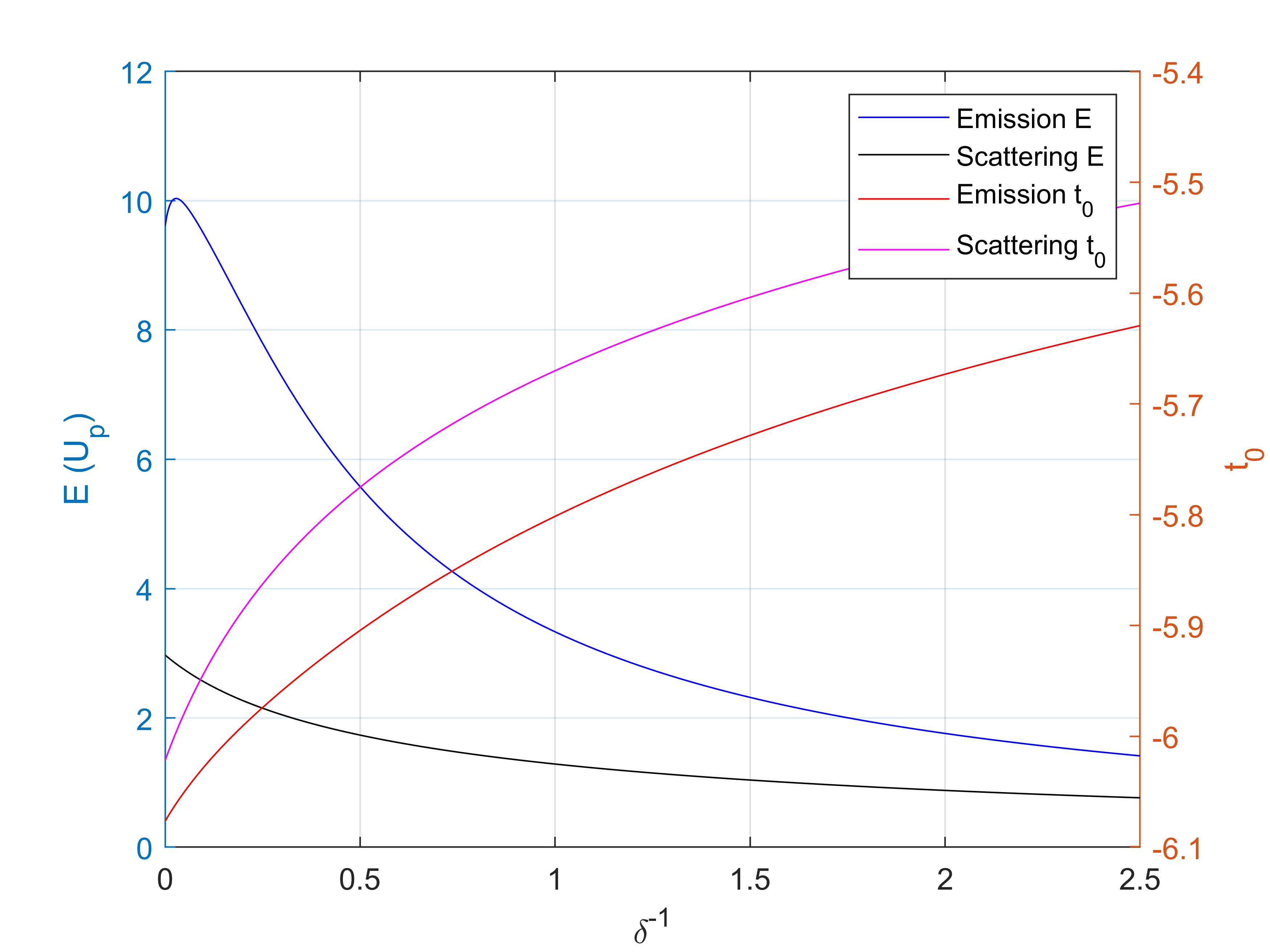}
	\caption{Cutoff behavior for pulsed ($\tau=6\pi$, $\phi=0$) laser using the exponential field profile. Emission (blue) and scattering (black) cutoff energies (left axis) are shown as a function of the adiabaticity parameter. Additionally, the times of ionization (right axis) that leads to these peak energies are plotted for the emission (red) and scattering (magenta) cutoffs.}
	\label{fig8}
\end{figure}

We may also find the peak energies as a function of ponderomotive amplitude just as before, however we will want to search in the three most likely regions: around $t_0=-4\pi$, $-2\pi$, and $0$ in Figure \ref{fig7}. The cutoff energies for this pulsed laser, using the exponential field profile, is shown in Figure \ref{fig8}. One may notice interesting behavior for very small ponderomotive amplitudes (large $\delta$) where the peak electron energy increases slightly before following the typical decay. This is the complexity we avoided by including the remaining ponderomotive energy at the end of the simulation. However, since the fields decay in time, we need not include this factor (as it would be effectively zero).

This interesting behavior can be attributed to the integrated ponderomotive force during the laser pulse following scattering. As ponderomotive amplitude increases, the electron will experience more of the laser field's profile and therefore gain some energy through the ponderomotive force. However, once the ponderomotive amplitude is sufficiently large, the weaker near- to mid-field takes part in the rescattering process and reduces the peak achievable energy. The idea that this effect occurs post-scattering is why we see no such structure in the scattering energy curve. As one may expect this initial rise in energy is extended for shorter pulses and vanishes for long pulses.

\subsection{CEP Dependence}

\begin{figure}[H]
	\centering
	\includegraphics[width=1\linewidth]{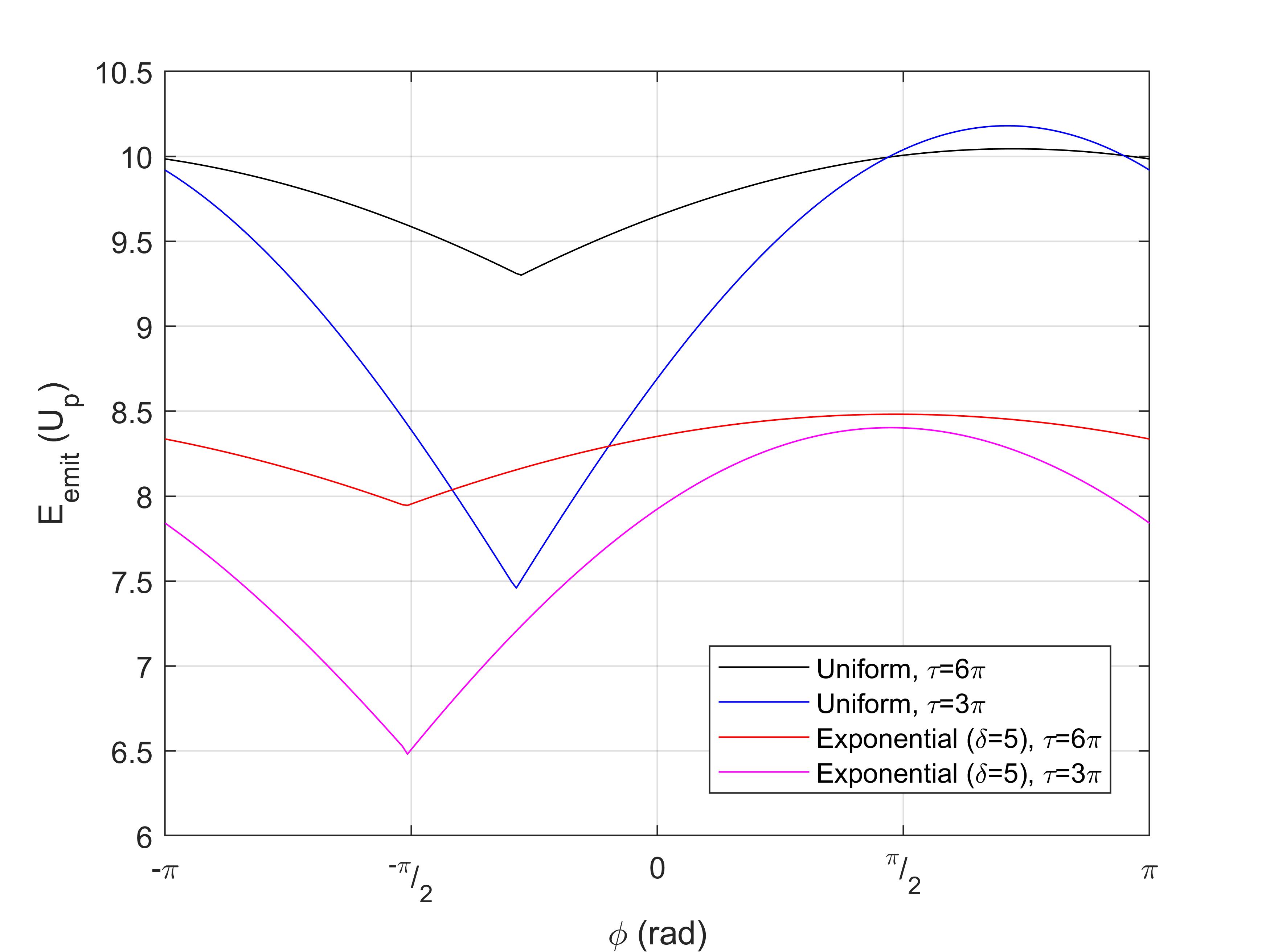}
	\caption{Electron emission classical cutoff as a function of CEP for systems with either few- or single-cycle pulses ($\tau=6\pi$ or $3\pi$) and for uniform or exponential field profiles.}
	\label{fig9}
\end{figure}

\begin{figure}[H]
	\centering
	\includegraphics[width=1\linewidth]{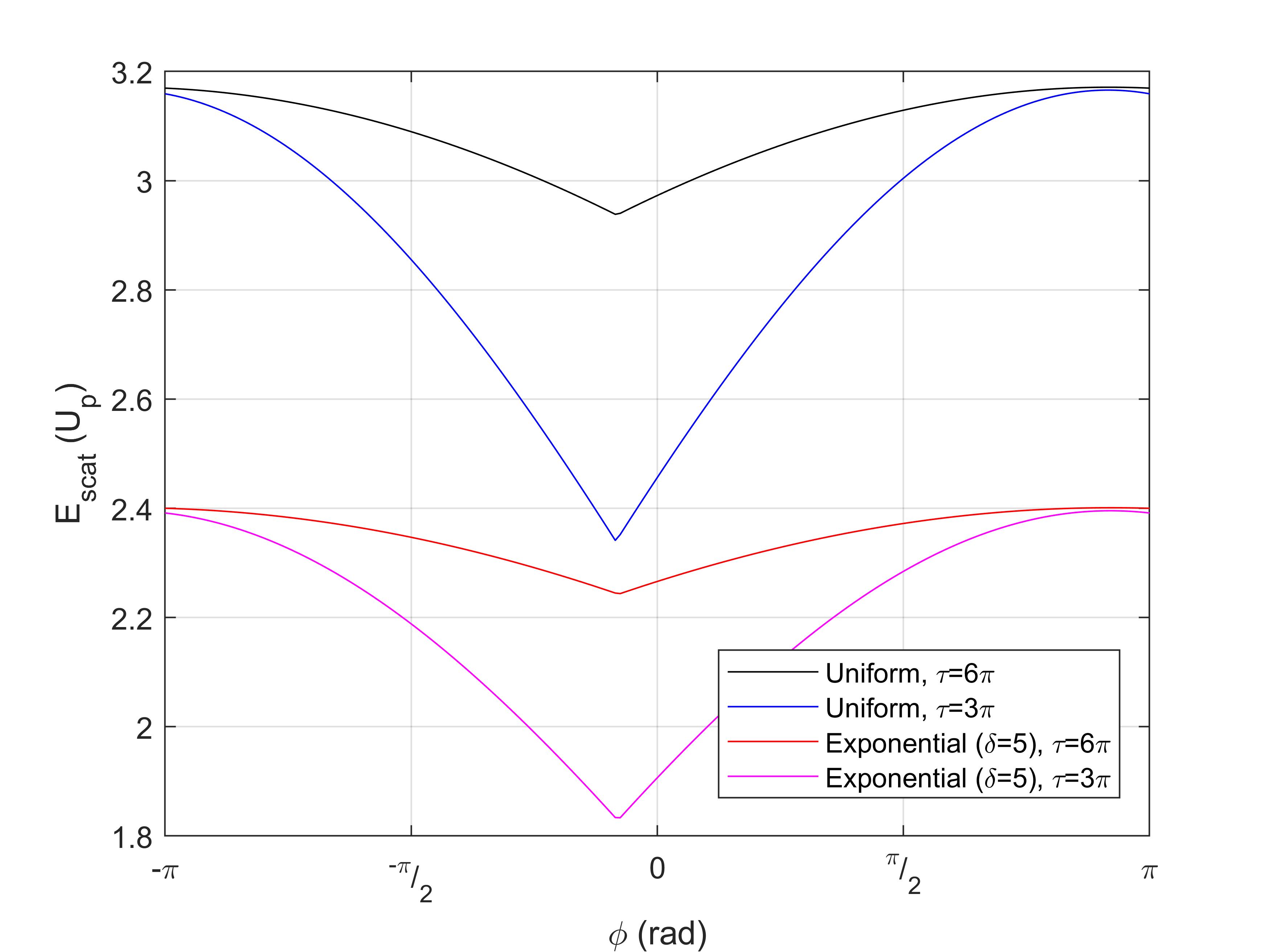}
	\caption{Scattering energy cutoff as a function of CEP for systems with either few- or single-cycle pulses ($\tau=6\pi$ or $3\pi$) and for uniform or exponential field profiles.}
	\label{fig10}
\end{figure}

To preface the next section investigating the effects of the CEP of a pulsed ionization laser, we found that adjusting results for the envelope-reduced peak field (by multiplying $E_0$ by $m(\phi)=\max_t\left|e^{-2\ln 2 \left(\frac{t}{\tau}\right)^2}\cos(t-\phi)\right|$) does not account for the phenomena we observe. This correction makes the proceeding curves (Figures \ref{fig9} and \ref{fig10}) roughly piecewise linear, but adds complexity due to the two differential discontinuities of $m(\phi)$ with respect to the CEP. Correcting for either peak extraction or peak restoring field only also makes the plots roughly linear, but does not aide in interpretation and gives rise to energies much larger than expected ($\sim 15 U_p$). For these reasons we continue with the original laser profile in Equation \ref{eq22}, recreating some of the results in Ref \cite{atcont}.

The emission cutoff behavior is shown in Figure \ref{fig9}. The uniform field results are equivalent to Figure 3a in Ref \cite{atcont} (although the CEP is offset by $\pi$ relative to our definition). As expected the dependence on CEP is strengthened with a shorter pulse and including a field gradient decreases the peak energy. For both uniform and exponential fields the peak energy tends to occur for $\phi\in[0, \pi]$, where $\phi=\pi/2$ corresponds to a sine-like pulse where peak extraction follows peak restoring fields and $\phi=\pi$ corresponds to a peak restoration force at $t=0$. This may unveil a choice of CEP targeting peak electron energy, which may be generally dependent on the geometry and fields at play. Additionally, the peak here is at about $10.18 U_p$ for the short pulse with a uniform field. Note that while the ponderomotive quantities are calculated for a peak field of 1, this field may not be achieved provided that $\phi\neq n\pi$ ($m(\phi)<1$). The sharp vertex in each of these curves indicates that the peak emission time transitioned from one cycle to another.

The CEP's effect on the scattering cutoff energy is shown in Figure \ref{fig10}. The phase for peak energy is consistently and expectedly near to $\phi=\pm\pi$ where the peak field at $t=0$ is restorative. Like the emission case, here we see an exaggerated effect for short pulses, and a general drop in photon energy when including a field gradient.

\section{Conclusion}

In this paper we have reproduced the semi-classical uniform field simulations originally used to show that the peak emission and scattering energies in the process of electron rescattering are $\sim10U_p$ and $\sim3.17U_p$ respectively.

We then proceeded to modify this calculation by applying spatial field profiles which were inspired by common nanostructures in experiments. The semi-classical energy limit for emission and scattering in these systems are provided as functions of the adiabaticity parameter which encodes how the field profile drops off in the near-field. Comparing the exponential profile to the tip and blade profiles revealed its theoretical limitations. For strong field drop-offs, $\delta\sim1$ or less, the model disagrees by up to nearly $0.5 U_p$ for tips and nearly $0.9 U_p$ for blades, indicating that higher orders of the field profile beyond $\delta$ are required to describe these strong field phenomena precisely.

We additionally applied a temporal profile to the laser field and observed a temporally sensitive ponderomotive force which may apply a slight boost to the already diminished peak emission energy. This boost is quickly quenched by the observed drop in peak energy associated with smaller $\delta$. We finally observed how the CEP affects the emission and scattering energies. The CEP becomes expectedly more important for shorter pulses. A choice of CEP may be made to maximize or minimize the cutoff energies for short pulses.

While our results for the electron emission energy should be accurate for most systems, the scattering energy is more dubious when applied to solids. Solid HHG involves the dynamics of the ionized electron and its associated electron hole within the material, and the scattering energy is the energy at which these two recombine. The inclusion of the electron hole, which would have a different effective mass, dispersion relation, and/or field profile, obsoletes the natural units of the system. Inclusion of these properties are important and have been done \cite{otherpap, classhhg}, although the results are less transferable between systems. Additionally, in cases where the ionization density is large, these single-body calculations may not be applicable \cite{singlim}.

\section{Acknowledgements}

This research was funded by the Center for Bright Beams, National Science Foundation Grant No. PHY-1549132.

\printbibliography

\end{document}